\documentclass[twocolumn,showkeys,aps,prb,showpacs]{revtex4-1}
\usepackage{graphicx}
\usepackage[CJKbookmarks,dvipdfm,colorlinks,linkcolor=blue,citecolor=blue]{hyperref}

\begin{document}

\title{Coexistence of intrinsic piezoelectricity and nontrivial band topology  in monolayer  InXO (X=Se and Te) }

\author{San-Dong Guo$^{1}$, Wen-Qi Mu$^{1}$, Yu-Tong Zhu$^{1}$, Shao-Qing Wang$^{1}$ and Guang-Zhao Wang$^{2}$}
\affiliation{$^1$School of Electronic Engineering, Xi'an University of Posts and Telecommunications, Xi'an 710121, China}

\affiliation{$^2$Key Laboratory of Extraordinary Bond Engineering and Advanced Materials Technology of Chongqing, School of Electronic Information Engineering, Yangtze Normal University, Chongqing 408100, China}
\begin{abstract}
The combination of piezoelectricity with other unique properties (like  topological
insulating phase and intrinsic ferromagnetism) in two-dimensional (2D) materials  is much worthy of intensive
study. In this work, the  piezoelectric properties of  2D
topological insulators  InXO (X=Se and Te) from monolayer InX (X=Se and Te) with double-side oxygen functionalization are studied by
density functional theory (DFT). The large  piezoelectric strain coefficients (e.g. $d_{11}$=-13.02 pm/V for InSeO and  $d_{11}$=-9.64 pm/V for InTeO) are predicted, which are comparable and even higher than ones of many  other  familiar 2D materials.
Moreover, we propose two strategies to enhance  piezoelectric response  of  monolayer  InXO (X=Se and Te).
Firstly, the biaxial   strain (0.94-1.06) is applied, and  the $d_{11}$ (absolute value) is  increased  by 53\%/56\% for monolayer InSeO/InTeO at 1.06 strain, which is due to increased $e_{11}$ (absolute value) and reduced $C_{11}-C_{12}$. In considered strain range, InXO (X=Se and Te) monolayers are  always 2D topological insulators, which confirm the coexistence of piezoelectricity and nontrivial band topology.
Secondly, a Janus monolayer $\mathrm{In_2SeTeO_2}$ is  designed by replacing the top Se/Te atomic layer in monolayer InSeO/InTeO with Te/Se atoms,  which is  dynamically  and mechanically stable.  More excitingly, Janus  monolayer $\mathrm{In_2SeTeO_2}$ is also a 2D topological insulator with sizeable bulk gap up to 0.158 eV,  confirming the coexistence of intrinsic piezoelectricity and topological nature.
The calculated $d_{11}$ is -9.9 pm/V, which is in the middle of ones of InSeO and InTeO monolayers.
Finally, the  carrier  mobilities of monolayer  InXO (X=Se and Te) are obtained, which  shows a rather pronounced anisotropy between electron and hole, and  are almost isotropic between armchair and zigzag directions.
Our works imply that it is possible to use the piezotronic effect to control the quantum  transport process, ultimately
leading to novel device applications in monolayer  InXO (X=Se and Te), and  can stimulate further experimental works.

\end{abstract}

\pacs{71.20.-b, 77.65.-j, 72.15.Jf, 78.67.-n ~~~~~~~~~~~~~~~~~~~~~~~~~~~~~~~~~~~Email:sandongyuwang@163.com}

\maketitle

\section{Introduction}
2D materials  can show a variety of  extraordinary new physical properties, like 2D piezoelectricity and quantum
spin Hall (QSH) phase.  The  piezoelectricity allows for energy conversion
between electrical and mechanical energy, and a 2D material should break inversion symmetry,
and then can exhibit piezoelectricity, which  has
attracted growing interest due to potential application in sensors, actuators and
energy sources\cite{q4}.
The monolayer $\mathrm{MoS_2}$ with 2H phase is  predicted as a typical 2D piezotronic material\cite{q11}, and then is  proved to possess piezoelectricity experimentally with $e_{11}$=2.9$\times$$10^{-10}$ C/m\cite{q5,q6}.
In theory, the piezoelectric properties of  many 2D materials  have been reported by DFT calculations\cite{q11,q7,q9,q10,q12,q13,q14,q15,q16,q17}, like  transition metal dichalchogenides (TMD), Janus TMD, group IIA and IIB metal oxides, group III-V semiconductors,  $\mathrm{MA_2Z_4}$ family, and  group-III monochalcogenides.

 Topological insulators  have rich physics and
promising applications in spintronics and quantum computations\cite{t1,t2}.
The QSH insulators are a novel quantum state, which can be  characterized by the gapless edge states inside the bulk gap. The charge carriers from edge states are robust against backscattering, which is very useful for  energy-efficient electronic devices.
The QSH insulators are  known as 2D topological insulators, which  is firstly proposed in graphene\cite{t3}, and  are experimentally verified in HgTe/CdTe and InAs/GaSb quantum wells\cite{t4,t5}. Many QSH
insulators have been theoretically proposed\cite{t6,t7,t8,t9,t10,cxq}, such as silicene,
Bi(111) bilayer, chemically modified Sn,  $\mathrm{ZrTe_5}$/ $\mathrm{HfTe_5}$,  $\mathrm{Bi_4Br_4}$ and $\mathrm{SrGa_2Se_4}$ .

\begin{figure*}
  \includegraphics[width=12.0cm]{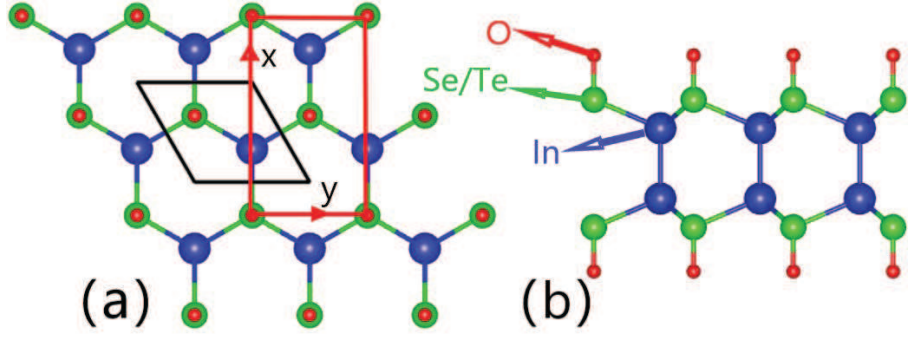}
  \caption{(Color online)The  crystal structure of monolayer InXO (X=Se and Te): top view (a) and side view (b).  The  rhombus primitive cell  and the rectangle supercell are shown by  black and red frames.}\label{t0}
\end{figure*}

\begin{figure}
  \includegraphics[width=8cm]{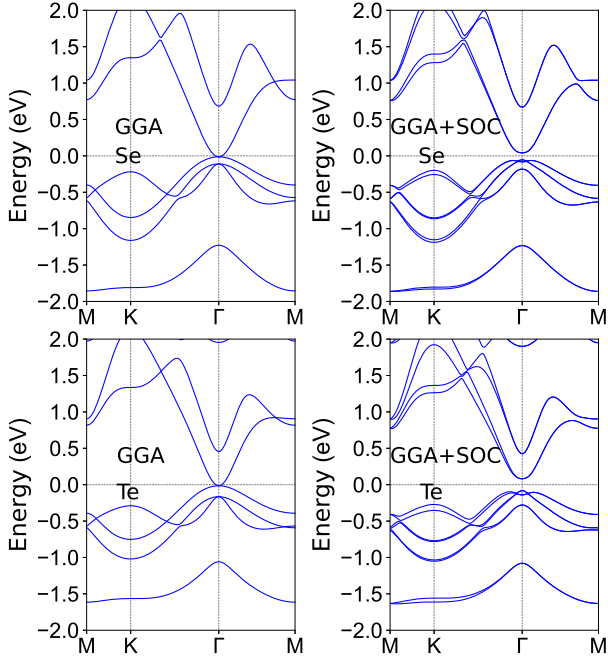}
\caption{(Color online)The energy band structures  of   monolayer InXO (X=Se and Te)  using GGA  and GGA+SOC.  }\label{band}
\end{figure}

Compared with 2D materials with individual  piezoelectric or
QSH characteristics,  2D piezoelectric
topological insulators  with  both piezoelectric and
QSH characteristics  will open up
unprecedented opportunities for intriguing physics, whose exploitation will ultimately
lead to novel device applications.  The combination of piezoelectricity and ferromagnetism has been achieved in 2D vanadium dichalcogenides and septuple-atomic-layer  $\mathrm{VSi_2P_4}$\cite{qt1,q15}, and a 2D ferroelastic topological insulator that simultaneously possesses ferroelastic and QSH characteristics has also been predicted in 2D Janus TMD MSSe (M = Mo and W)\cite{qt2}.
However, to the best of our knowledge, no studies have been reported on  combination of piezoelectricity and QSH insulators.

The  group-III monochalcogenides with broken inversion symmetry have piezoelectricity\cite{q16}, and the  piezoelectricity of  Janus group-III chalcogenide monolayers  can be enhanced
with respect to perfect group-III monochalcogenide.
monolayers\cite{q17}.
Recently, monolayer group-III monochalcogenides  by oxygen
functionalization  are predicted as a promising class of 2D topological insulators\cite{qt3}, which break inversion symmetry, allowing  these materials to become piezoelectric. In this work,  the piezoelectric properties of monolayer  InXO (X=Se and Te)  are reported by using DFT.
The large $d_{11}$ of monolayer  InXO (X=Se and Te) are predicted, and they are -13.02 pm/V for InSeO and -9.64 pm/V for InTeO.
The biaxial strain is  used to tune their piezoelectric properties, and  the improvement is by 53\%/56\% for monolayer InSeO/InTeO at 1.06 strain.
Inspiring from experimentally synthesized MoSSe monolayer\cite{msse},  a Janus $\mathrm{In_2SeTeO_2}$ monolayer with dynamical and mechanical stability  is  designed, and it is a 2D topological insulator with the gap value of 0.158 eV. The predicted $d_{11}$ is -9.99 pm/V, which falls between ones of InSeO and InTeO monolayers. In all our studied systems,  the coexistence of piezoelectricity and nontrivial band topology is confirmed, which may provide a new platform for intriguing physics and novel device applications.

The rest of the paper is organized as follows. In the next
section, we shall give our computational details and methods.
 In  the next few sections,  we shall present crystal and electronic structures, piezoelectric properties, strain and Janus effects on piezoelectric properties, and carrier mobilities of monolayer  InXO (X=Se and Te).  Finally, we shall give our discussion and conclusions.

\begin{figure}
  \includegraphics[width=8cm]{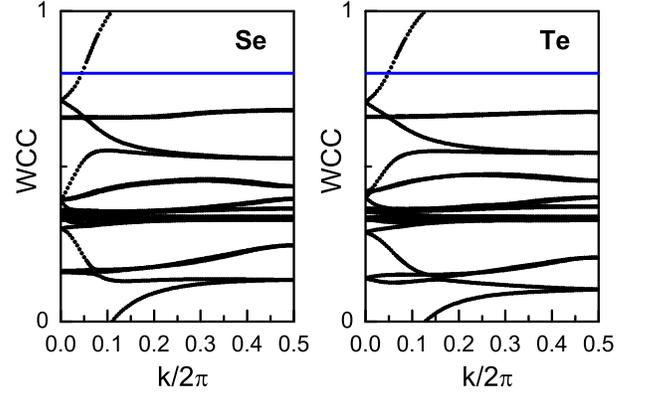}
\caption{(Color online)Evolution of the WCC of monolayer InXO (X=Se and Te), confirming the topological nature of the gap with
$Z_2$=1. }\label{z2}
\end{figure}

\begin{table}
\centering \caption{For InSeO (InTeO) monolayer, the lattice constants $a_0$ ($\mathrm{{\AA}}$), the elastic constants $C_{ij}$ ($\mathrm{Nm^{-1}}$), shear modulus
$G_{2D}$ ($\mathrm{Nm^{-1}}$),  Young's modulus $C_{2D}$  ($\mathrm{Nm^{-1}}$),  Poisson's ratio $\nu$,  the GGA+SOC gaps  (eV) and  $Z_2$ topological invariant. }\label{tab0}
  \begin{tabular*}{0.48\textwidth}{@{\extracolsep{\fill}}cccc}
  \hline\hline
$a_0$& $C_{11}$/$C_{22}$ &  $C_{12}$& $G_{2D}$\\\hline
4.46 (4.78) &27.13 (24.85)&11.21 (9.57)&7.96 (7.64)\\\hline\hline
$C_{2D}$& $\nu$& $Gap$& $Z_2$\\\hline
22.50 (21.17)&0.41 (0.39)&0.096 (0.163) & 1(1)\\\hline\hline
\end{tabular*}
\end{table}
\begin{figure*}
   \includegraphics[width=15.6cm]{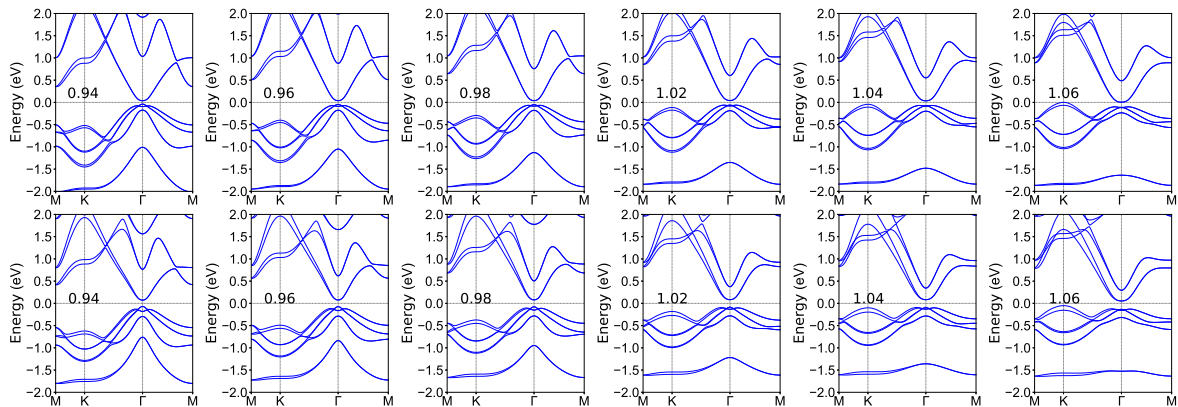}
\caption{(Color online) The energy band structures  of monolayers InSeO (Top) and InTeO (Bottom)  using GGA+SOC with $a/a_0$ changing from 0.94 to 1.06.}\label{band-s}
\end{figure*}
\begin{figure}
  \includegraphics[width=8cm]{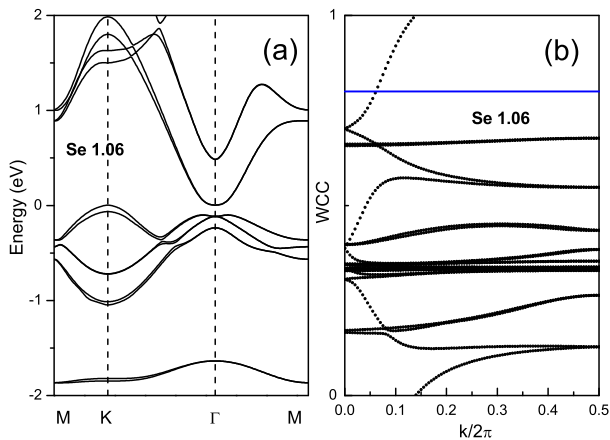}
\caption{(Color online) (a) The  fitted energy band of InSeO by Wannier90 and (b)Evolution of WCC of  InSeO at 1.06 strain. }\label{z2-1}
\end{figure}
\begin{figure*}
   \includegraphics[width=12cm]{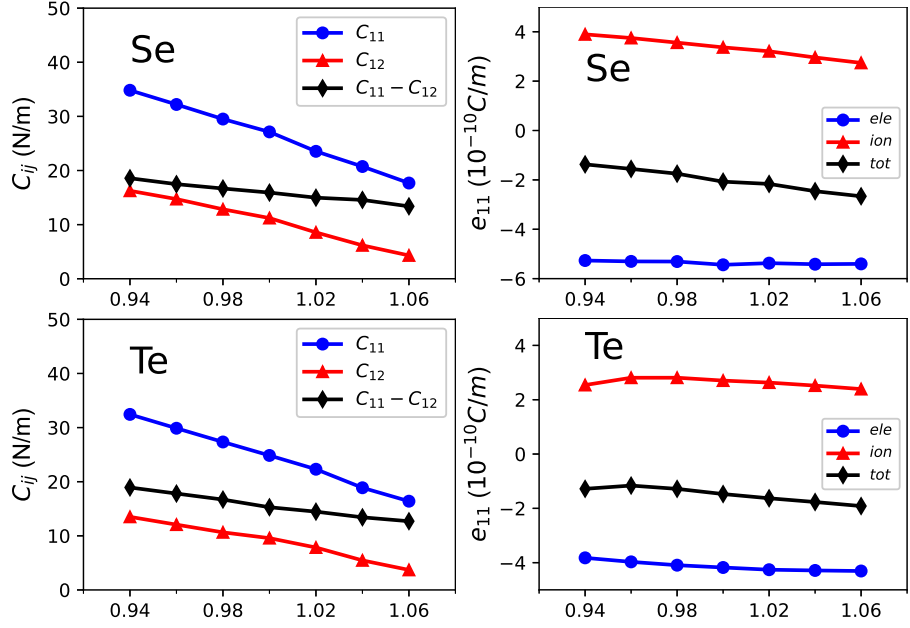}
  \caption{(Color online) For monolayer  InXO (X=Se and Te),  the elastic constants  $C_{ij}$ and  piezoelectric stress coefficients  $e_{11}$ along with  the ionic  and electronic contributions to  $e_{11}$,  with the application of  biaxial strain (0.94 to 1.06).}\label{s}
\end{figure*}

\section{Computational detail}
The first-principles calculations within DFT\cite{1} are performed using  the projected augmented wave
(PAW) method, as implemented in
the  VASP package\cite{pv1,pv2,pv3}.   The cutoff energy
for plane-wave expansion is 500 eV, and the vacuum
region along the z direction is set to about 20 $\mathrm{{\AA}}$ in order to
to avoid interactions
between two neighboring images.
The popular generalized gradient approximation of Perdew, Burke and  Ernzerhof  (GGA-PBE)\cite{pbe} is used as the exchange-correlation potential. The spin-orbit coupling (SOC) is included to study electronic structures and piezoelectric stress coefficients $e_{ij}$.
 K-point meshes of
16$\times$16$\times$1  are employed for geometry optimization and self-consistent electronic structure calculations.
The total energy  convergence criterion is set
to $10^{-8}$ eV. The geometry optimization was considered
to be converged with the residual force on each atom being less than 0.0001 $\mathrm{eV.{\AA}^{-1}}$.

To obtain the piezoelectric strain coefficients $d_{ij}$, the elastic stiffness tensor $C_{ij}$ are calculated by using  strain-stress relationship (SSR), and the piezoelectric stress coefficients $e_{ij}$ are calculated by density functional perturbation theory (DFPT) method\cite{pv6}.
The 2D elastic coefficients $C^{2D}_{ij}$
 and   piezoelectric stress coefficients $e^{2D}_{ij}$
have been renormalized by the the length of unit cell along z direction ($Lz$):  $C^{2D}_{ij}$=$Lz$$C^{3D}_{ij}$ and $e^{2D}_{ij}$=$Lz$$e^{3D}_{ij}$.
The Brillouin zone sampling
is done using a Monkhorst-Pack mesh of 16$\times$16$\times$1  for $C_{ij}$, and  5$\times$10$\times$1 for $e_{ij}$.
The $Z_2$ invariants are calculated by the Wannier90 and WannierTools
codes\cite{w1,w2}, where a tight-binding Hamiltonian with the
maximally localized Wannier functions is fitted to the first-principles band structures.
We use Phonopy code\cite{pv5} to calculate  phonon dispersion
spectrums of $\mathrm{In_2SeTeO_2}$ monolayer with a supercell
of 5$\times$5$\times$1. The finite displacement method is adopted, and  a 2$\times$2$\times$1 k-mesh is employed with kinetic energy cutoff of 500 eV.

\begin{table*}
\centering \caption{Piezoelectric coefficients $e_{11}(d_{11})$ of  monolayer InSe, InSeO, InTe and InTeO, along with $C_{ij}$. The unit is $10^{-10}$C/m,   pm/V and $\mathrm{Nm^{-1}}$ for $e_{ij}$, $d_{ij}$ and $C_{ij}$. The previous available theoretical values are given in parentheses. }\label{tab-y}
  \begin{tabular*}{0.96\textwidth}{@{\extracolsep{\fill}}cccccc}
  \hline\hline
Name & $e_{11}$ &  $C_{11}$&$C_{12}$&$C_{11}-C_{12}$&$d_{11}$\\\hline\hline
InSe&-0.69 (-0.57\cite{q16}, -0.70\cite{q17})& 49.74 (51\cite{q16}, 55\cite{q17})&13.55(12\cite{q16}, 20\cite{q17})&36.20 &-1.92 (-1.46\cite{q16}, -1.98\cite{q17})\\\hline
InSeO& -2.07&    27.13   &11.21 & 15.92&-13.02                               \\\hline\hline
InTe&-0.53(-0.69\cite{q17})& 41.23 (64\cite{q17})&10.43(6\cite{q17}) & 30.80 &-1.73 (-1.18\cite{q17})\\\hline
InTeO&-1.47& 24.85& 9.57 & 15.28 &-9.64 \\\hline\hline
\end{tabular*}
\end{table*}

\begin{table}
\centering \caption{The electronic (Ele) and   ionic (Ion) contributions to $e_{11}$ of  monolayer InSe, InSeO, InTe and InTeO. }\label{tab-cc}
  \begin{tabular*}{0.48\textwidth}{@{\extracolsep{\fill}}ccccc}
  \hline\hline
Name & InSe &InSeO& InTe&InTeO\\\hline\hline
Ele&-5.14 & -5.44 &   -4.55&    -4.18                       \\\hline
Ion& 4.44  & 3.37&     4.02 &    2.71                 \\\hline\hline
\end{tabular*}
\end{table}
\section{Crystal and electronic structures}
The crystal structure of  monolayer  InXO (X=Se and Te) can be constructed from monolayer InX (X=Se and Te)  by
oxygen functionalization with chemisorption of oxygen atoms on  both sides.  Monolayer InX (X=Se and Te) has a hexagonal lattice, which is
composed of two In atomic layers sandwiched between X atomic layers.
The geometric structure of monolayer  InXO (X=Se and Te) is shown in \autoref{t0}, and both rhombus primitive cell  and rectangle supercell are plotted. The rectangle supercell is used to calculate carrier mobilities and piezoelectric coefficients with armchair and zigzag directions  be defined as x and y directions. The  monolayer InXO (X=Se and Te) has the same point group $\bar{6}m2$ with g-$\mathrm{C_3N_4}$ and $\mathrm{MoS_2}$\cite{gsd}.
The optimized lattice constant of monolayer InSeO and InTeO are 4.46  $\mathrm{{\AA}}$ and  4.78 $\mathrm{{\AA}}$, and the bond
length between O and X atoms is 1.67 $\mathrm{{\AA}}$ and 1.84 $\mathrm{{\AA}}$, which agree well previous theoretical values\cite{qt3}.
It is found that the lattice constants of  monolayer  InXO are  larger than ones of monolayer InX (InSe:4.09 $\mathrm{{\AA}}$ and InTe:4.38 $\mathrm{{\AA}}$) due to  oxygen functionalization.

In ref.\cite{qt3}, it is proved that monolayer  InXO (X=Se and Te) possess  energetic, thermal and chemical stabilities. Here, we further check
the mechanical stability of the monolayer  InXO (X=Se and Te) by
elastic constants $C_{ij}$. Using Voigt notation, the elastic tensor can be expressed as:
\begin{equation}\label{pe1-4}
   C=\left(
    \begin{array}{ccc}
      C_{11} & C_{12} & 0 \\
     C_{12} & C_{11} &0 \\
      0 & 0 & (C_{11}-C_{12})/2 \\
    \end{array}
  \right)
\end{equation}
The  two  independent elastic
constants of monolayer InSeO/InTeO are $C_{11}$=27.13 $\mathrm{Nm^{-1}}$/24.85 $\mathrm{Nm^{-1}}$ and $C_{12}$=11.21 $\mathrm{Nm^{-1}}$/9.57 $\mathrm{Nm^{-1}}$.   The $C_{66}$ can be attained by ($C_{11}$-$C_{12}$)/2, and they are 7.96 $\mathrm{Nm^{-1}}$/7.64 $\mathrm{Nm^{-1}}$, which are also  shear modulus $G^{2D}$. The mechanical stability of a material with hexagonal symmetry  should satisfy the  Born  criteria\cite{ela}:
 \begin{equation}\label{e1}
C_{11}>0,~~ C_{66}>0
\end{equation}
 The calculated $C_{11}$ and  $C_{66}$ confirm  the mechanical stability of monolayer  InXO (X=Se and Te).
The Young's modulus $C_{2D}(\theta)$ are given\cite{ela1}:
\begin{equation}\label{c2d}
C_{2D}(\theta)=\frac{C_{11}C_{22}-C_{12}^2}{C_{11}sin^4\theta+Asin^2\theta cos^2\theta+C_{22}cos^4\theta}
\end{equation}
where $A=(C_{11}C_{22}-C_{12}^2)/C_{66}-2C_{12}$. Due to  hexagonal symmetry, the  monolayer  InXO (X=Se and Te) are  mechanically isotropic.
 The calculated $C_{2D}$ is 22.50 $\mathrm{Nm^{-1}}$/21.17 $\mathrm{Nm^{-1}}$ for  InSeO/InTeO monolayer.
 These values are obviously smaller than those of other 2D materials\cite{q11,q7,q9,q10,q12,q13,q14,q15,q16,q17}, which
means that InSeO and InTeO monolayers are more flexible than other 2D materials.
 The Poisson's ratio $\nu(\theta)$ is also isotropic, and can be expressed as:
 \begin{equation}\label{e1}
\nu^{2D}=\frac{C_{12}}{C_{11}}
\end{equation}
 The calculated  $\nu$ is 0.41/0.39 for InSeO/InTeO monolayer.

The energy bands of monolayer  InXO (X=Se and Te) with GGA and GGA+SOC are shown in \autoref{band}.
The GGA results show that the monolayer  InXO (X=Se and Te) are semimetals with the valence band maximum (VBM) and conduction band minimum (CBM)
degenerated at the $\Gamma$ point. When the SOC is included, a direct gap of 0.096 eV/0.163 eV is observed for InSeO/InTeO monolayer.
 The transition from
semimetal to insulator induced by SOC  suggests that InXO (X=Se and Te) monolayers are potential 2D topological insulators.
 In order to ascertain the topological phase transition
in the monolayer  InXO (X=Se and Te), we calculate the $Z_2$
topological invariants with  $Z_2$ = 1 being a topologically nontrivial state and  $Z_2$ = 0 being a trivial state.
Since the  InXO (X=Se and Te) monolayers lack
inversion symmetry, the $Z_2$ invariants cannot be determined from the parities of the filled states.
However, the topologically nontrivial nature  can be confirmed via calculations of the Wannier charge center (WCC), as
plotted in \autoref{z2}. It is clearly seen that the number of crossings
between the WCC and the reference horizontal line is odd, which confirms  the topological nature of monolayer  InXO (X=Se and Te) with
$Z_2$ = 1.

\section{Piezoelectric properties}
 The pristine InX (X=Se and Te) monolayers are piezoelectric\cite{q16}, and  their piezoelectric effect can be enhanced by designing Janus structures\cite{q17}.
 When the  InX (X=Se and Te) monolayers with both sides are fully covered by O atoms [InXO (X=Se and Te) monolayers], they have the same  point group symmetry with InX monolayers, which is because  the O-X bonds are perpendicular to the InX layers. Therefore, InXO (X=Se and Te) monolayers are also piezoelectric. The piezoelectric effects of a material can be described by  third-rank piezoelectric stress tensor  $e_{ijk}$ and strain tensor $d_{ijk}$ from the sum of ionic
and electronic contributions,  which are defined as:
 \begin{equation}\label{pe0}
      e_{ijk}=\frac{\partial P_i}{\partial \varepsilon_{jk}}=e_{ijk}^{elc}+e_{ijk}^{ion}
 \end{equation}
and
 \begin{equation}\label{pe0-1}
   d_{ijk}=\frac{\partial P_i}{\partial \sigma_{jk}}=d_{ijk}^{elc}+d_{ijk}^{ion}
 \end{equation}
In which $P_i$, $\varepsilon_{jk}$ and $\sigma_{jk}$ are polarization vector, strain and stress, respectively.
The $e_{ijk}^{elc}$/$d_{ijk}^{elc}$ is clamped-ion piezoelectric coefficients, while  the  $e_{ijk}$/$d_{ijk}$ is relax-ion piezoelectric coefficients as a realistic result.

For 2D materials, $\varepsilon_{jk}$=$\sigma_{ij}$=0 for i=3 or j=3\cite{q9,q10,q11,q12}.
 Due to a $\bar{6}m2$ point-group symmetry of InXO (X=Se and Te) monolayers,
 the  piezoelectric stress   and strain tensors with Voigt notation can be reduced into:
 \begin{equation}\label{pe1-1}
 e=\left(
    \begin{array}{ccc}
      e_{11} & -e_{11} & 0 \\
     0 & 0 & -e_{11} \\
      0 & 0 & 0 \\
    \end{array}
  \right)
    \end{equation}

  \begin{equation}\label{pe1-2}
  d= \left(
    \begin{array}{ccc}
      d_{11} & -d_{11} & 0 \\
      0 & 0 & -2d_{11} \\
      0 & 0 &0 \\
    \end{array}
  \right)
\end{equation}
where $e_{11}$/$d_{11}$  represents the in-plane  piezoelectric
stress/strain component,  which  is induced  by
uniaxial in-plane  strain. The $e_{11}$ can be calculated by DFPT, and the
$d_{11}$  can be attained by the relation:
 \begin{equation}\label{pe1-3}
    e=dC
 \end{equation}
 Here, the  $d_{11}$  is derived by  \autoref{pe1-1}, \autoref{pe1-2}, \autoref{pe1-3} and \autoref{pe1-4}:
\begin{equation}\label{pe2-7}
    d_{11}=\frac{e_{11}}{C_{11}-C_{12}}
\end{equation}

\begin{figure}
   \includegraphics[width=7.0cm]{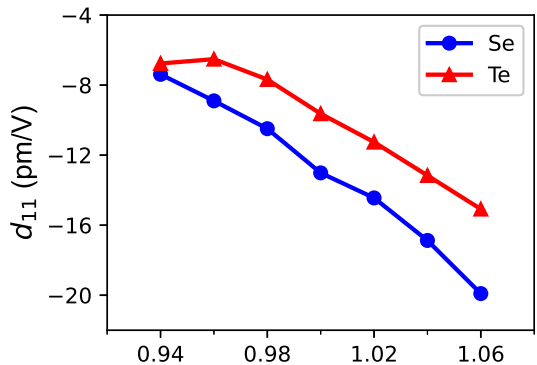}
  \caption{(Color online) For monolayer  InXO (X=Se and Te), the piezoelectric strain coefficients  $d_{11}$  with the application of  biaxial strain (0.94 to 1.06).}\label{s1}
\end{figure}

The  orthorhombic supercell as the computational unit cell is used to calculate  $e_{ij}$ by DFPT.
The calculated in-plane $e_{11}$ ($d_{11}$) of  monolayer  InSeO and InTeO are -2.07$\times$$10^{-10}$ C/m (-13.02 pm/V) and   -1.47$\times$$10^{-10}$ C/m (-9.64 pm/V).
The   $d_{11}$ (absolute values) of  monolayer  InXO (X=Se and Te) is much higher than those of the widely used bulk piezoelectric materials\cite{aln,aln-1,aln-2,aln-3} such as  $\alpha$-quartz ($d_{11}$=2.3 pm/V),  wurtzite-AlN ($d_{33}$=5.1 pm/V) and wurtzite-GaN ($d_{33}$=3.1 pm/V). Their $d_{11}$ are also comparable to or even superior than ones of many familiar 2D materials\cite{q7,q9,q11,q12,q7-2,q7-1}, such as $\mathrm{MoS_2}$ ($d_{11}$=3.65 pm/V), $\mathrm{WS_2}$ ($d_{11}$=2.12 pm/V), ZnO ($d_{11}$=8.65 pm/V), MoSSe ($d_{11}$=3.76 pm/V) and MoSTe ($d_{11}$=5.04pm/V).
It is found that  the $e_{11}$ (absolute values) of monolayer  InXO (X=Se and Te) is
comparable to or even smaller than ones of  $\mathrm{MoS_2}$ ($e_{11}$=3.62$\times$$10^{-10}$ C/m ), $\mathrm{WS_2}$ ($e_{11}$=2.43$\times$$10^{-10}$ C/m ), ZnO ($e_{11}$=2.66$\times$$10^{-10}$ C/m ), MoSSe ($e_{11}$=3.74$\times$$10^{-10}$ C/m ) and MoSTe ($e_{11}$=4.53$\times$$10^{-10}$ C/m )\cite{q7,q9,q7-2,q7-1}, but their
$d_{11}$ are larger than  ones of those monolayers.
The  underlying reason is that the  $G_{2D}$ of monolayer  InXO (X=Se and Te) is very smaller than ones of $\mathrm{MoS_2}$ (49.7 $\mathrm{Nm^{-1}}$)\cite{q9}, $\mathrm{WS_2}$ (57.2 $\mathrm{Nm^{-1}}$)\cite{q9}, ZnO (15.4 $\mathrm{Nm^{-1}}$)\cite{q9}, MoSSe (49.7 $\mathrm{Nm^{-1}}$)\cite{q7} and WSTe (45 $\mathrm{Nm^{-1}}$)\cite{q7},  which leads to larger $d_{11}$  based on \autoref{pe2-7} (The $d_{11}$ is inversely proportional to $G_{2D}$.).
These results  show that monolayer  InXO (X=Se and Te) may have large in-plane piezoelectric response, when  a uniaxial strain is applied.

 To  consider oxygen functionalization effects on piezoelectric properties,  the piezoelectric coefficients  of pristine InX (X=Se and Te) monolayers also are calculated, and their $C_{ij}$, $e_{11}$ and $d_{11}$  along with previous theoretical values are listed in \autoref{tab-y}. Our calculated results agree with previous available ones\cite{q16,q17}. Upon double-side functionalization, the $e_{11}$ are improved  as well to -2.07$\times$$10^{-10}$ C/m/-1.47$\times$$10^{-10}$ C/m from -0.69$\times$$10^{-10}$ C/m/-0.53$\times$$10^{-10}$ C/m for monolayer InSeO/InTeO, and the $C_{11}-C_{12}$ are reduced to 15.92 $\mathrm{Nm^{-1}}$/15.28 $\mathrm{Nm^{-1}}$ from 36.20 $\mathrm{Nm^{-1}}$/30.80 $\mathrm{Nm^{-1}}$. These give rise to enhanced piezoelectric effect by oxygen functionalization. Finally, the ionic  and electronic contributions to $e_{11}$ of   monolayer  InXO  and InX (X=Se and Te) are listed in \autoref{tab-cc}.
It is clearly seen that the ionic  and electronic contributions are opposite for all four monolayers, and the electronic contribution dominates the $e_{11}$. Compared to monolayer InX(X=Se and Te), the oxygen functionalization can lead to small difference between the ionic  and electronic contributions for monolayer InXO (X=Se and Te), which induces the larger $e_{11}$. Thus, it is very important for calculating $e_{11}$ to use relaxed-ion, not clamped-ion.

\begin{table*}
\centering \caption{The calculated  elastic modulus ($C_{2D}$), effective mass ($m^*$), deformation potential ($E_l$) and carrier mobility ($\mu_{2D}$) [300 K]   of InSeO and InTeO monolayers.}\label{tab-u2d}
  \begin{tabular*}{0.96\textwidth}{@{\extracolsep{\fill}}ccccccc}
  \hline\hline
&Carrier type&    &$C_{2D}$ ($\mathrm{Nm^{-1}}$) & $m^*$ ($m_0$) & $E_l$ (eV)& $\mu_{2D}$ ($\mathrm{cm^2V^{-1}s^{-1}}$)\\\hline\hline
&Electrons   & x&  22.50&       0.799&       -8.56         & 10.21    \\
InSeO&            &y&     22.50&       0.804 &      -8.57         & 10.13                                           \\
&Holes   & x&        22.50&      -0.428&     -8.59         &35.50                                              \\
&             &y&    22.50&      -0.427&     -8.58           &35.66                                 \\\hline\hline
&Electrons   & x&  21.17&        0.538&       -8.47       & 21.38  \\
InTeO&            &y&    21.17&       0.555&      -8.47          & 20.73                                            \\
&Holes   & x&       21.17&      -0.245&     -8.54           &100.77                                              \\
&             &y&   21.17&      -0.256&     -8.53          &96.66                              \\\hline\hline
\end{tabular*}
\end{table*}

\begin{figure}
  \includegraphics[width=8cm]{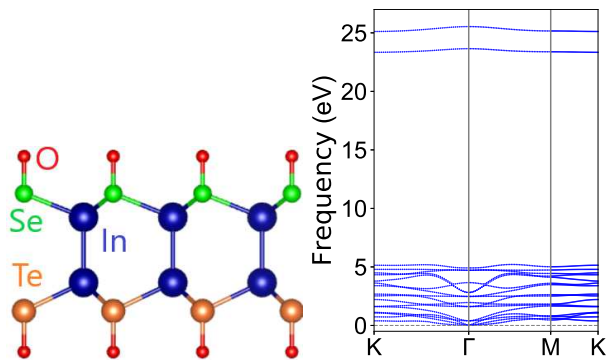}
\caption{(Color online)Left:The side view of crystal structure of monolayer $\mathrm{In_2SeTeO_2}$; Right: the phonon band dispersions of monolayer $\mathrm{In_2SeTeO_2}$. }\label{se-s}
\end{figure}

\begin{figure}
  \includegraphics[width=8cm]{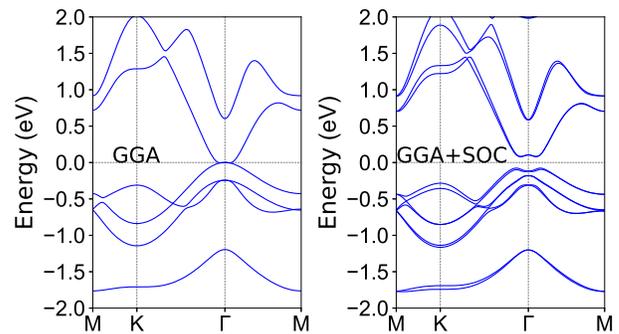}
\caption{(Color online)The energy band structures  of   monolayer $\mathrm{In_2SeTeO_2}$  using GGA  and GGA+SOC.  }\label{band-sete}
\end{figure}

\section{Strain effects}
Strain is a very effective method to tune  piezoelectric properties of 2D materials\cite{gsd,s1,s2,s3}, and the $d_{11}$ of monolayer $\mathrm{MoS_2}$ and g-$\mathrm{C_3N_4}$ with the same space group (No. 187) of  monolayer InXO (X=Se and Te) can be improved by biaxial tensile strain.
Compared to unstrained one, tensile strain of  4\% can  raise  the $d_{11}$ of $\mathrm{MoS_2}$/g-$\mathrm{C_3N_4}$  by about 46\%/330\%\cite{gsd}. So, it is very interesting to investigate biaxial strain effects on piezoelectric properties  of monolayer InXO (X=Se and Te).  For space group (No. 187), the biaxial strain can not produce polarization, but the uniaxial strain along x or y direction can induce polarization charges.
Here, we only consider  biaxial strain ($a/a_0$ from 0.94 to 1.06) effects on  piezoelectric properties of monolayer  InXO (X=Se and Te).

At applied strain, the monolayer InXO (X=Se and Te)  should be a semiconductor to exhibit piezoelectricity, and  the strain-related energy bands  are plotted in \autoref{band-s}. It is clearly seen that monolayer InXO (X=Se and Te) during the whole strain range all  have a gap, and
the SOC gap never closes,  which means that the nontrivial topology is
robust under considered compressive/tensile strains. To  precisely confirm coexistence of intrinsic piezoelectricity and nontrivial band topology,
the $Z_2$ topological invariants are calculated for all considered ones, and they  all satisfy   $Z_2$=1. Here, we only show the  fitted energy band of  monolayer InSeO by Wannier90 and evolution of the WCC of  monolayer InSeO at 1.06 strain in \autoref{z2-1}. These  mean that the piezoelectricity can coexist with nontrivial band topology in all considered strained monolayer InXO (X=Se and Te).

The  elastic constants $C_{ij}$, piezoelectric coefficients $e_{11}$ along  the ionic  and electronic contributions, and  $d_{11}$ of  monolayer InXO (X=Se and Te)  as a function of strain
are plotted in \autoref{s} and \autoref{s1}.
With the  strain  from 0.94 to 1.06, all  $C_{ij}$  monotonously decrease, and  the reduced $C_{11}-C_{12}$  is in favour of improving  $d_{11}$ according to \autoref{pe2-7}.
It is found that  piezoelectric coefficient $e_{11}$ (absolute values)  increases with strain from 0.94 to 1.06, which can give rise to enhanced $d_{11}$. At 1.06 strain, the $d_{11}$ of InSeO/InTeO  reaches up to -19.91 pm/V/-15.08 pm/V from unstrained -13.02 pm/V/-9.64  pm/V, increased by 53\%/56\%. Thus, tensile strain can improve  piezoelectric response of monolayer InXO (X=Se and Te).

\section{Janus structure}
By constructing the Janus structure, the piezoelectric response ($d_{11}$) of the pristine InX (X=Se and Te) monolayers can be enhanced\cite{q17}.
The $d_{11}$  of $\mathrm{In_2SeTe}$ can be improved to 4.73 pm/V from 1.98 pm/V of InSe or  1.18 pm/V of InTe\cite{q17}.
It's a natural idea to achieve Janus 2D monolayer from monolayer InXO (X=Se and Te).  To
construct the Janus structure, the  $\mathrm{In_2SeTeO_2}$  monolayer can be achieved by replacing the top Se/Te atomic layer in monolayer InSeO/InTeO with Te/Se atoms. The symmetry of monolayer $\mathrm{In_2SeTeO_2}$ (No.156) is lower than that of the monolayer InXO (X=Se and Te) (No.187) due to
the lack of  the reflection symmetry with respect to the central  In atomic bilayer. The reduced symmetry can induce many novel properties, like out-of-plane  piezoelectric polarizations.

The side view of crystal structure of monolayer $\mathrm{In_2SeTeO_2}$ is shown in \autoref{se-s}, along with its phonon band dispersions.
The optimize lattice constants are $a$=$b$=4.63 $\mathrm{{\AA}}$, which is between ones of InSeO and InTeO.
The bond lengths of  monolayer $\mathrm{In_2SeTeO_2}$ between O and Se/Te atoms is 1.67 $\mathrm{{\AA}}$/1.84 $\mathrm{{\AA}}$, which is the same with one of monolayer InSeO/InTeO.
The phonon dispersions show no imaginary bands, signifying
its dynamic stability. The out-of-plane vibration of the O atoms  are at frequency of 23 eV-26 eV, and there are two phonon bands due to inequitable O atoms of both sides, which is different from one of InSeO or InTeO\cite{qt3}. The other phonon bands are below 5.3 eV.
The calculated elastic constants satisfy the  Born  criteria of mechanical stability with $C_{11}$=25.85 $\mathrm{Nm^{-1}}$ and $C_{12}$=10.46 $\mathrm{Nm^{-1}}$. The other elastic physical quantities are calculated, such as $G_{2D}$=7.69 $\mathrm{Nm^{-1}}$, $C_{2D}$=21.62 $\mathrm{Nm^{-1}}$ and $\nu$=0.41.

\begin{figure}
  \includegraphics[width=8cm]{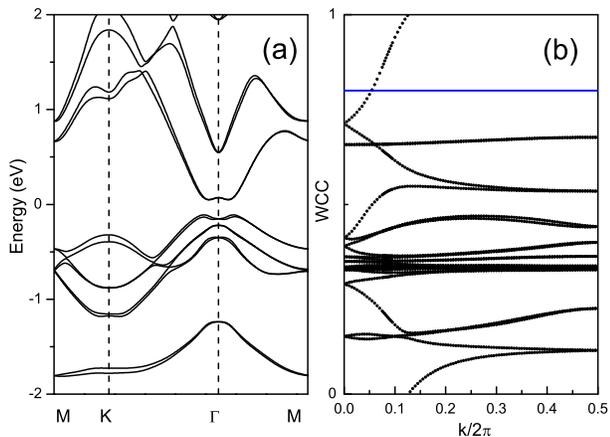}
\caption{(Color online) (a) The  fitted energy band of monolayer $\mathrm{In_2SeTeO_2}$ by Wannier90 and (b)Evolution of WCC of  monolayer $\mathrm{In_2SeTeO_2}$. }\label{z2-sete}
\end{figure}

\begin{figure}
   \includegraphics[width=8cm]{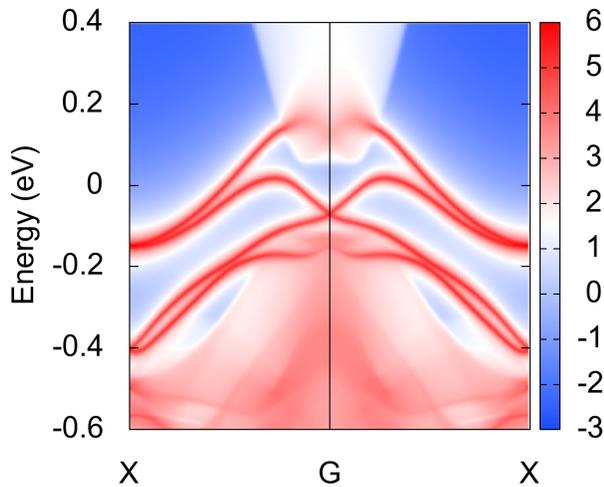}
\caption{(Color online) Topological edge states connecting the conduction and valence bands of monolayer $\mathrm{In_2SeTeO_2}$ with a band gap of
0.158 eV.}\label{z2-s}
\end{figure}

The energy band structures  of   monolayer $\mathrm{In_2SeTeO_2}$  using GGA  and GGA+SOC are shown in \autoref{band-sete}. The SOC can induce that monolayer $\mathrm{In_2SeTeO_2}$ changes  from
semimetal to insulator (the gap of 0.158 eV), which  suggests that it may be  a  potential  topological insulator. To confirm the topologically nontrivial character of the
gap, we further evaluate the $Z_2$ invariant by tracing the evolution of the
WCC, which is shown in \autoref{z2-sete} along with the  fitted energy band of monolayer $\mathrm{In_2SeTeO_2}$ by Wannier90. The $Z_2$  invariant is identified by the odd number crossings of WCC (black lines) of the reference line (blue line), and the $Z_2$=1, which indicates that the monolayer $\mathrm{In_2SeTeO_2}$ is a 2D  topological insulator, possessing the helical edge states connecting the conduction
and valence bands.  And, we use Green's-function method to calculate the surface states on (100) surface based on the tight-binding
Hamiltonian, which is plotted in \autoref{z2-s}. Clearly, the edge states of monolayer $\mathrm{In_2SeTeO_2}$ are present in the energy gap.

The monolayer InXO (X=Se and Te) [$\bar{6}m2$ symmetry]
possess a reflection symmetry with respect to  the central  In atomic bilayer, which  requires that $e_{31}$/$d_{31}$=0. In
monolayer $\mathrm{In_2SeTeO_2}$, the  inequivalent Se-In and Te-In bonding lengths break the reflection symmetry along the
vertical direction, resulting in a low degree of $3m$ symmetry. And then, both $e_{11}$/$d_{11}$ and $e_{31}$/$d_{31}$ are nonzero.
The piezoelectric stress   and strain tensors will become  into:
 \begin{equation}\label{pe1-1-1}
 e=\left(
    \begin{array}{ccc}
      e_{11} & -e_{11} & 0 \\
     0 & 0 & -e_{11} \\
      e_{31} & e_{31} & 0 \\
    \end{array}
  \right)
    \end{equation}

  \begin{equation}\label{pe1-2-2}
  d= \left(
    \begin{array}{ccc}
      d_{11} & -d_{11} & 0 \\
      0 & 0 & -2d_{11} \\
      d_{31} & d_{31} &0 \\
    \end{array}
  \right)
\end{equation}
According to above discussions, when  a uniaxial in-plane strain is applied, both monolayer InXO (X=Se and Te) and  $\mathrm{In_2SeTeO_2}$
have  in-plane piezoelectric polarization, but monolayer $\mathrm{In_2SeTeO_2}$ has
 an additional vertical piezoelectric polarization.
When they are subject to biaxial in-plane strain,  the
in-plane piezoelectric polarization will be suppressed for both monolayer InXO (X=Se and Te) and  $\mathrm{In_2SeTeO_2}$, while the
out-of-plane one still will remain in monolayer $\mathrm{In_2SeTeO_2}$. Based on  \autoref{pe1-1-1}, \autoref{pe1-2-2}, \autoref{pe1-3} and \autoref{pe1-4}, the $d_{11}$ and $d_{31}$ can be expressed as:
\begin{equation}\label{pe2-2}
    d_{11}=\frac{e_{11}}{C_{11}-C_{12}}~~~and~~~d_{31}=\frac{e_{31}}{C_{11}+C_{12}}
\end{equation}

The calculated $e_{11}$ of  monolayer $\mathrm{In_2SeTeO_2}$  is -1.54$\times$$10^{-10}$ C/m from the the ionic contribution 2.98$\times$$10^{-10}$ C/m  and electronic contribution -4.52$\times$$10^{-10}$ C/m. Based on \autoref{pe2-2}, the calculated $d_{11}$ is -9.99 pm/V, which  falls between ones of the InSeO and InTeO. Similar phenomenon can be found in TMD and $\mathrm{MA_2Z_4}$ families\cite{q7,gsd1}. For example, the $d_{11}$ of the Janus $\mathrm{MSiGeN_4}$ (M=Mo and W) monolayers are between those of the
$\mathrm{MSi_2N_4}$ (M=Mo and W) and $\mathrm{MGe_2N_4}$ (M=Mo and W) monolayers\cite{gsd1}.
The predicted $d_{31}$ is up to 26.25 pm/V due to very large $e_{31}$ of 9.53 $\times$$10^{-10}$ C/m, which is mainly from ionic contribution (93\%).  The  more accurate calculation with  a 7$\times$12$\times$1 k-mesh  confirms the very large $d_{31}$.

\section{Carrier mobility}
Finally, we also investigate the carrier mobilities of  electron/hole of  monolayer InXO (X=Se and Te) by the deformation potential (DP) theory  proposed by Bardeen and Shockley\cite{dp}, and the carrier
mobility of a 2D material  ($\mu_{2D}$)  can be expressed as:
\begin{equation}\label{u2d}
  \mu_{2D}=\frac{e\hbar^3C_{2D}}{K_BTm^*m_dE_l^2}
\end{equation}
where  the electron charge, the reduced Planck
constant and  the Boltzmann constant are marked by $e$, $\hbar$ and $K_B$, and the  temperature,  effective mass in the transport direction and the average effective mass are shown by the  $T$ and   $m^*$ and $m_d=\sqrt{m_xm_y}$. In addition, $E_l$ represents the DP constant, as defined by $E_l=\Delta E/\delta$, where $\Delta E$ can be calculated by the band edge of CBM or VBM minus the vacuum level. The strain range is chose from -0.01 to 0.01 with the step $\delta$=0.005.

The rectangular supercell is used to calculate the  carrier mobilities of  monolayer InXO (X=Se and Te) with the temperature being   300 K.  Firstly, the effective masses of CBM and VBM along x and y directions are calculated by:
\begin{equation}
(m_{l})^{-1}=\frac{1}{\hbar^2}\frac{\partial^2E(k)}{\partial k_l^2}
\end{equation}
where  $E(k)$ is the dispersion of the lowest conduction band/the highest valence band. To fit second-order polynomial $E=(\hbar k)^2/2m$,
 the 4 k points  with a spacing of
0.003 reciprocal lattice constants are used to produce the inverse mass
tensor. It is found that the  effective masses of holes of monolayer InXO (X=Se and Te) are smaller than ones of electrons, and the effective masses between x and y directions for both electrons and holes are very close.
 The DP constant $E_l$ can be attained by linearly  fitting the band energies of the VBM and CBM  with respect to the vacuum energy, and the related slopes are  DP constant $E_l$. Calculated results show that all $E_l$ of  both the VBM and CBM of  both monolayer InSeO and InTeO almost are the same, and about -8.5.

 Based on \autoref{u2d}, the carrier mobilities of monolayer InXO (X=Se and Te) are attained, which are shown \autoref{tab-u2d}.
The hole mobilities of monolayer InXO (X=Se and Te) are in the range of 35.50-100.77
$\mathrm{cm^2V^{-1}s^{-1}}$, which are larger than those of
electrons  (10.13-21.38 $\mathrm{cm^2V^{-1}s^{-1}}$). This is because the holes have smaller effective masses than electrons.
It is found that the
carrier mobilities of  both electrons and holes between x and y directions show very weak anisotropy,
which  is because the effective masses and $E_l$ between x and y directions are very close.

\section{Discussions and Conclusion}
Due to expensive computation time to calculate $C_{ij}$ and $e_{ij}$ using hybrid functional HSE06 or HSE06+SOC, we use GGA or GGA+SOC to
investigate the $C_{ij}$ and $e_{ij}$ of monolayer  InXO (X=Se and Te).
Although the GGA  may  underestimate semiconductor gaps of  monolayer  InXO (X=Se and Te), our conclusion should be  qualitatively correct.
The intrinsic piezoelectricity and nontrivial band topology can coexist in Janus monolayer $\mathrm{In_2SeTeO_2}$. In fact, Janus monolayer $\mathrm{In_2SSeO_2}$ and  $\mathrm{In_2STeO_2}$ are all topological insulators, and they can also achieve coexistence of intrinsic piezoelectricity and nontrivial band topology.

In summary, we demonstrate piezoelectric QSH insulators in monolayer  InXO (X=Se and Te) by using first-principles calculations.
The  biaxial strain and constructing Janus structure are used to tune  piezoelectric properties of monolayer  InXO (X=Se and Te).
It is found that tensile strain can effectively improve the $d_{11}$ of monolayer  InXO (X=Se and Te), and the strained systems are all QSH insulators.
The Janus monolayer $\mathrm{In_2SeTeO_2}$ is predicted to be a QSH insulator with sizable nontrivial band gap,  which is dynamically and mechanically stable. The calculated $d_{11}$ of monolayer $\mathrm{In_2SeTeO_2}$ is between ones of monolayer InSeO and InTeO.
The unique coexistence of piezoelectricity and QSH insulator in  monolayer  InXO (X=Se and Te) along with $\mathrm{In_2SeTeO_2}$ offers the opportunities for achieving multi-function control electronic devices.
Our work will inspire further researches to explore  piezoelectric QSH insulators.

\begin{acknowledgments}
This work is supported by Natural Science Basis Research Plan in Shaanxi Province of China  (2019JQ-860). We are grateful to the Advanced Analysis and Computation Center of China University of Mining and Technology (CUMT) for the award of CPU hours and WIEN2k/VASP software to accomplish this work.
\end{acknowledgments}


\begin{references}

\bibitem{q4}W. Wu and Z. L. Wang, Nat. Rev. Mater. \textbf{1}, 16031 (2016).

\bibitem{q11}K. N. Duerloo, M. T. Ong and E. J. Reed, J. Phys. Chem. Lett. \textbf{3}, 2871 (2012).


\bibitem{q5} W. Wu, L. Wang, Y. Li, F. Zhang, L. Lin, S. Niu, D. Chenet,
X. Zhang, Y. Hao, T. F. Heinz, J. Hone and Z. L. Wang,
Nature \textbf{514}, 470 (2014).


\bibitem{q6}H. Zhu, Y. Wang, J. Xiao, M. Liu, S. Xiong, Z. J. Wong, Z. Ye,
Y. Ye, X. Yin and X. Zhang, Nat. Nanotechnol. \textbf{10},
151 (2015).

\bibitem{q7}L. Dong, J. Lou and V. B. Shenoy, ACS Nano, \textbf{11},
8242 (2017).




\bibitem{q9}M. N. Blonsky, H. L. Zhuang, A. K. Singh and R.  G. Hennig,  ACS Nano, \textbf{9},
9885 (2015).

\bibitem{q10}R. X. Fei, We. B. Li, J. Li and L. Yang, Appl. Phys. Lett.  \textbf{107}, 173104 (2015).




\bibitem{q12}Y. Chen,  J. Y. Liu,  J. B. Yu,  Y. G. Guo and Q. Sun, Phys. Chem. Chem. Phys.
 \textbf{21}, 1207 (2019).


\bibitem{q13} S. D. Guo, Y. T. Zhu, W. Q. Mu and W. C. Ren,  EPL \textbf{132},  57002 (2020).


\bibitem{q14}S. D. Guo, Y. T. Zhu, W. Q. Mu, L. Wang  and  X. Q. Chen,   Comp. Mater. Sci. \textbf{188}, 110223 (2021)

\bibitem{q15}S. D. Guo, W. Q. Mu, Y. T. Zhu and X. Q. Chen, Phys. Chem. Chem. Phys. \textbf{22}, 28359 (2020).

\bibitem{q16}W. B. Li  and J. Li, Nano Res.  \textbf{8}, 3796 (2015).

\bibitem{q17}Y. Guo, S. Zhou, Y. Z. Bai, and J. J. Zhao, Appl. Phys. Lett. \textbf{110}, 163102 (2017).


\bibitem{t1}M. Z. Hasan and C. L. Kane, Rev. Mod. Phys. Kane, \textbf{82}, 3045 (2010).


\bibitem{t2}X. L. Qi and  S. C. Zhang, Rev. Mod. Phys. \textbf{83}, 1057 (2011).


\bibitem{t3}C. L. Kane and  E. J.  Mele, Phys. Rev. Lett. \textbf{95}, 226801 (2005).


\bibitem{t4}M. Konig,   S. Wiedmann,   C. Brune et al., Science \textbf{318}, 766 (2007).


\bibitem{t5}I. Knez,  R. R. Du and  G. Sullivan, Phys. Rev. Lett. \textbf{107},
136603 (2011).

\bibitem{t6}C. C. Liu,   W. Feng,  Y.  Yao, Phys. Rev. Lett.  \textbf{107}, 076802 (2011).


\bibitem{t7}S. Murakami, Phys. Rev. Lett. \textbf{97}, 236805 (2006).

\bibitem{t8}Y. Xu,   B. Yan,  H. J. Zhang et al., Phys. Rev. Lett. \textbf{111}, 136804 (2013).

\bibitem{t9}H. M. Weng,  X. Dai and Z. Fang,  Phys. Rev. X  \textbf{4}, 011002 (2014).

\bibitem{t10}J. J. Zhou, W. X. Feng, C. C. Liu, S. Guan and Y. G. Yao, Nano Lett. \textbf{14}, 4767 (2014).
\bibitem{cxq}L. Wang,  Y. P.  Shi,  M. F.  Liu et al.,   arXiv:2008.02981 (2020).

\bibitem{qt1}J. H. Yang,  A. P. Wang, S. Z. Zhang, J.  Liu, Z. C. Zhong and L. Chen, Phys. Chem. Chem. Phys.,
\textbf{21}, 132 (2019).

\bibitem{qt2}Y. D. Ma, L. Z. Kou, B. B. Huang, Y. Dai and T. Heine, Phys. Rev. B \textbf{98}, 085420 (2018).


\bibitem{qt3}S. Zhou, C. C. Liu, J. J. Zhao and Y. G. Yao, npj Quant. Mater. \textbf{3}, 16 (2018).

\bibitem{msse}A. Y. Lu, H. Y. Zhu, J. Xiao et al., Nature Nanotechnology \textbf{12}, 744 (2017).
\bibitem{1}P. Hohenberg and W. Kohn, Phys. Rev. \textbf{136},
B864 (1964); W. Kohn and L. J. Sham, Phys. Rev. \textbf{140},
A1133 (1965).

\bibitem{pv1} G. Kresse, J. Non-Cryst. Solids \textbf{193}, 222 (1995).

\bibitem{pv2} G. Kresse and J. Furthm$\ddot{u}$ller, Comput. Mater. Sci. 6, \textbf{15} (1996).

\bibitem{pv3} G. Kresse and D. Joubert, Phys. Rev. B \textbf{59}, 1758 (1999).

\bibitem{pbe}J. P. Perdew, K. Burke and M. Ernzerhof, Phys. Rev. Lett. \textbf{77}, 3865 (1996).


 \bibitem{pv6}X. Wu, D. Vanderbilt and  D. R.  Hamann, Phys. Rev. B  \textbf{72}, 035105 (2005).


\bibitem{w1} Q. Wu, S. Zhang, H. F. Song, M. Troyer and A. A. Soluyanov, Comput. Phys. Commun. \textbf{224}, 405
(2018).
\bibitem{w2}A. A. Mostofia, J. R. Yatesb, G. Pizzif, Y.-S. Lee, I. Souzad, D.
Vanderbilte and N. Marzarif,  Comput. Phys. Commun. \textbf{185}, 2309 (2014).


\bibitem{pv5}A. Togo, F. Oba, and I. Tanaka, Phys. Rev. B \textbf{78}, 134106
(2008).



\bibitem{gsd}S. D. Guo, W. Q. Mu and Y. T. Zhu, J. Phys. Chem. Solids \textbf{151}, 109896 (2021).


\bibitem{ela}R. C. Andrew, R. E. Mapasha, A. M. Ukpong and N. Chetty, Phys. Rev. B \textbf{85}, 125428 (2012).
\bibitem{ela1}E. Cadelano, P. L. Palla, S. Giordano and L. Colombo,  Phys. Rev. B  \textbf{82}, 235414 (2010).
\bibitem{aln}K. Tsubouchi and N. Mikoshiba, IEEE Trans. Sonics Ultrason. \textbf{SU-32},
 634 (1985).

\bibitem{aln-1} C. M. Lueng, H. L. Chang, C. Suya and C. L. Choy, J. Appl. Phys. \textbf{88},
 5360 (2000).

\bibitem{aln-2} A. Hangleiter, F. Htzel, S. Lahmann and U. Rossow,  Appl. Phys. Lett.
\textbf{83},  1169 (2003).

\bibitem{aln-3} S. Muensit, E. M. Goldys and I. L. Guy,  Appl. Phys. Lett. \textbf{75},
3965 (1999).




\bibitem{q7-2}S. D. Guo, X. S. Guo, R. Y. Han and Y. Deng, Phys. Chem.
Chem. Phys. \textbf{21}, 24620 (2019).


\bibitem{q7-1}M. Yagmurcukardes, C. Sevik and F. M. Peeters, Phys. Rev. B  \textbf{100}, 045415 (2019).


\bibitem{s1}N. Jena, Dimple, S. D.  Behere  and A. D. Sarkar, J. Phys. Chem. C  \textbf{121}, 9181 (2017).





\bibitem{s2}Dimple, N. Jena, A. Rawat, R.  Ahammed,
M. K. Mohanta and A. D. Sarkar, J. Mater. Chem. A  \textbf{6},  24885 (2018).


\bibitem{s3}S. D. Guo, X. S. Guo, Y. Y. Zhang and K. Luo, J. Alloy. Compd. \textbf{822}, 153577 (2020).

\bibitem{gsd1}S. D. Guo, W. Q. Mu, Y. T. Zhu, R. Y. Han and W. C. Ren, J. Mater. Chem. C, 2021,  DOI: 10.1039/D0TC05649A.

\bibitem{dp}S. Bruzzone and G. Fiori, Appl. Phys. Lett. \textbf{99}, 222108 (2011).
\end{references}
\end{document}